\documentclass[12pt]{iopart}
\usepackage{graphicx,amsfonts,amssymb,amsbsy}

\begin{document}

\title[Nucleation rate of color superconducting droplets in protoneutron stars]{Nucleation rate of color superconducting droplets in protoneutron stars}

\author{T. A. S. do Carmo}
\address{Universidade Federal do ABC, Rua Santa Ad\'elia, 166, 09210-170, Santo Andr\'e, Brazil.}

\author{ G. Lugones}
\address{Universidade Federal do ABC, Rua Santa Ad\'elia, 166, 09210-170, Santo Andr\'e, Brazil.}
\ead{german.lugones@ufabc.edu.br}

\author{A. G. Grunfeld}
\address{Department of Physics, Sultan Qaboos University, P.O.Box: 36 Al-Khode 123 Muscat, Sultanate of Oman.} 
\address{CONICET, Rivadavia 1917, (1033) Buenos Aires, Argentina.}
\address{Departamento de F\'\i sica, Comisi\'on Nacional de  Energ\'{\i}a At\'omica, (1429) Buenos Aires, Argentina.}

\begin{abstract}
We analyse the nucleation of quark matter droplets under protoneutron star conditions. We adopt a two-phase framework in which the hadronic phase is described through a non-linear Walecka model and the just deconfined matter by the MIT bag model including color superconductivity.  Surface tension and curvature energy are calculated self-consistently within the MRE formalism. We impose flavour conservation during the transition, which means that the just deconfined quark droplet is transiently out of equilibrium with respect to weak interactions. Our results show that trapped neutrinos slightly increase the critical density for deconfinement and that color superconductivity significantly decreases such density at low  temperatures. We also show that the nucleation rate is negligible for droplets larger than $100-200$ fm  and is huge for smaller droplets provided that the temperature is low enough. We compare our results with previous calculations using the Nambu-Jona-Lasinio model with color superconductivity and the MIT bag model without color superconductivity. We conclude that the deconfinement transition should be triggered instantaneously when a density slightly larger than the bulk transition density is reached at some layer of a protoneutron star. Since color superconductivity lowers the transition density at low temperatures, the transition is likely to occur after significant cooling in a massive enough protoneutron star.
\end{abstract}

\submitto{\JPG}
\maketitle

\section{Introduction}

Understanding the behaviour of the strong interacting matter for moderate temperatures and high baryonic density is one of the main tasks in astrophysics for describing the interior of neutron stars and regions close to the core of collapsing stars.  In the interior of that objects, at high densities, the deconfinement transition from hadronic to quark matter might occur. This transition, in astrophysical compact and dense objects, is basically studied in two scenarios: in protoneutron stars (PNS) and neutron stars (NS). The PNS are compact objects remaining after the gravitational collapse and supernovae explosion of a massive star. Initially the temperature is about few tens of MeV and the neutrinos are temporarily trapped in the interior of the PNS. After a few seconds of evolution, the neutrinos are radiated, and the PNS cools down turning into a NS. The interior of these objects can reach a density well above the nuclear saturation density ($\rho_0$), and in this scenario the hadronic matter can suffer a deconfinement transition into quark matter \cite{Logoteta, Fraga, Lugones_CarmoNJL, Lugones2009, Hempel}. The phase transition should begin with the nucleation of a small ($\sim 10^{-14}$ m) quark-matter droplet near the center of the star  \cite{Lugones_CarmoNJL,IidaSato1998,Lugones2005,Lugones9899,Madsen1994,Bombaci2004,Bombaci2007} that may later grow through a combustion process converting a macroscopic portion of the star into quark matter \cite{Lugones1994}.

Even though hadrons are a bound state of quarks, we can not describe with a single model the thermodynamics of the transition from hadronic to quark matter. Then, we need to adopt different models to describe each phase.
In the present paper we analyse the deconfinement transition
in protoneutron star conditions employing the MIT bag
model in the description of quark matter. For the hadronic phase
we use a model based on a relativistic Lagrangian of hadrons
interacting via the exchange of $\sigma$, $\rho$, and $\omega$ mesons. The nucleation is treated as a first order phase transition and the finite size effects  of creating a droplet of
deconfined quark matter in the hadronic environment are described using the multiple reflection expansion (MRE) framework \cite{Lugones_Grunfeld2011,Balian1970, Madsen, Kiriyama2003, Kiriyama2005}.
Two important features we consider in our work are: (a) quark flavor is conserved during the deconfinement transition because it is driven by strong interactions  \cite{IidaSato1998,Lugones2005,Lugones9899,Madsen1994,Bombaci2004,Bombaci2007} \footnote{Works that  deal with the structure of hybrid proto-neutron stars \cite{asked_by_referee},  consider the hadron-quark interphase as being in equilibrium under weak interactions. This condition is appropriate for such studies but it isn't for the present analysis of the deconfinement of quark droplets.} and (b) when color superconductivity is included together with flavor conservation, the most likely configuration of the just deconfined phase is two-flavor color superconductor (2SC) provided the pairing gap is large enough \cite{Lugones2005,Lugones_Grunfeld2011}.

The article is organized as follows. In Sec. 2 we present the main aspects of the non--linear Walecka model
describing the hadronic phase. In Sec. 3 we present the generalities of the model we use for the quark phase and the MRE formalism.  In Sec. 4 we study the deconfinement transition at finite temperature for different neutrino trapping conditions. In Sec. 5 we present our results followed by a {{summary}} and conclusions in Sec. 6.

\section{Hadronic matter}\label{HM}

We use a non-linear Walecka model for describing the hadronic phase (see e.g.\ \cite{Lugones_CarmoNJL} and references therein) composed by the baryon octet ($n$, $p$, $\Lambda$, $\Sigma^{+}$, $\Sigma^{0}$, $\Sigma^{-}$, $\Xi^{-}$, $\Xi^{0}$), electrons $e^{-}$, electron neutrinos $\nu_{e}$ and the corresponding antiparticles.  The Lagrangian for this model is the following \cite{Glendenning1991}
\begin{eqnarray}
{\cal L} & = & {\cal L}_{B} + {\cal L}_{M} + {\cal L}_{L}  \nonumber \\
& = & \sum_{B= n, p,  \Lambda, \Sigma^{+,0,-},  \Xi^{-,0}} \bar \psi_B  \bigg[\gamma^\mu  (i\partial_\mu -   x_{\omega B} \ g_{\omega}  \omega_\mu  - x_{\rho B} \ g_{\rho} \ \vec \tau \cdot \vec \rho_\mu )    \nonumber\\
& - &  (m_B - x_{\sigma B} \ g_{\sigma} \sigma)\bigg] \psi_B  + \frac{1}{2} (\partial_{\mu} \sigma \partial^{\mu} \sigma -m_\sigma^2  \sigma^2) - \frac{b}{3}  m_N (g_\sigma\sigma)^3 -\frac{c}{4}  (g_\sigma \sigma)^4  \nonumber\\ 
& - &   \frac{1}{4} \omega_{\mu\nu} \omega^{\mu\nu} +\frac{1}{2} m_\omega^2 \omega_{\mu}\ \omega^{\mu} -\frac{1}{4} \vec \rho_{\mu\nu} \cdot \vec \rho\ \! ^{\mu\nu}   + \frac{1}{2}\ m_\rho^2\  \vec \rho_\mu \cdot \vec \rho\ \! ^\mu   \nonumber\\
& + &  \sum_{L =e, \nu_e} \bar{\psi}_{L}    [ i \gamma_{\mu}  \partial^{\mu}  - m_{L} ]\psi_{L},
\label{octetlag}
\end{eqnarray}
where $B$, $M$ and $L$ refer to baryons, mesons and leptons respectively. Baryons interact through the exchange of mesons $\sigma$, $\omega$ and $\rho$. The constants $x_{\sigma B} = g_{\sigma B}/g_{\sigma}$,   $x_{\omega B} = g_{\omega B} / g_{\omega}$ and  $x_{\rho B} = g_{\rho B}/g_{\rho}$ are the ratios of the coupling constants of the hyperons to the coupling constants of the nucleons. The EoS is obtained from the above Lagrangian {{by means of}} a relativistic mean field model treatment (see  \cite{Lugones_CarmoNJL} and references therein for more details). In order to adapt the equation of state (EoS) to the actual conditions prevailing in protoneutron stars, the lepton term includes the contribution of electron neutrinos.

The thermodynamic potential for this model is:
\begin{eqnarray}
\Omega_H &=& -\sum_{B,L} \Omega_{i} -\frac{1}{2}\left(\frac{g_{\omega}}{m_{\omega}}\right)^{2}\rho_{B}^{'2}
+\frac{1}{2}\left(\frac{g_{\sigma}}{m_{\sigma}}\right)^{2}(g_{\sigma}\sigma)^{2} \nonumber\\
&&+ \frac{1}{3}bm_{n}(g_{\sigma}\sigma)^{3}  +\frac{1}{4}c(g_{\sigma}\sigma)^{4}-\frac{1}{2}\left(\frac{g_{\rho}}{m_{\rho}}\right)^{2}\rho_{I_{3}}^{'2}.
\label{omega_hadrons}
\end{eqnarray}
The weighted  isospin density  $\rho^{'}_{I_3}$ and  the weighted
baryon density $\rho^{'}_{B}$ are given by:
\begin{equation}
\rho^{'}_{I_3}  = \sum_{i=B}{ x_{\rho i} I_{3 i} n_i} ,
\end{equation}
\begin{equation}
\rho^{'}_{B}  = \sum_{i=B}{ x_{\omega i} n_i} ,
\end{equation}
where $I_{3 i}$ is the third component of the  isospin of each baryon and $n_i$ is the particle number density of each baryon:
\begin{equation}
n_i =   {\gamma_{i} \over{(2 \pi)^3} }
\int { d^{3}p   \;   ( f_i(T) - \bar{f}_i(T) )} ,
\label{num}
\end{equation}
being $\gamma_{i}$ the degeneracy factor.

The mean field $g_{\sigma} \sigma$ satisfies the equation:
\begin{equation}
\bigg({g_{\sigma}\over{m_{\sigma}}} \bigg)^{-2}
(g_{\sigma} \sigma)
+  b m_n (g_{\sigma} \sigma)^2
+  c (g_{\sigma} \sigma)^3
=  \sum_{i=B}{ x_{\sigma i} n^s_i} ,
\end{equation}
where $n^s_i$ is the scalar density:
\begin{equation}
n^s_i =
{\gamma_{i} \over{(2 \pi)^3} }
\int { d^{3}p
\;    { {m_i^{*}} \over{(p^2 + m_i^{*2})^{1/2}} }
\;     ( f_i(T) +  \bar{f}_i(T) )
} .
\end{equation}

From the thermodynamic potential we obtain the total pressure $P$  and the mass - energy  density $\rho$:
\begin{eqnarray}
P^H & = & \sum_{i=B,L}{ P_i }
+ {1\over{2}} \bigg({g_{\omega}\over{m_{\omega}}} \bigg)^2 \rho_{B}^{'2}
- {1\over{2}} \bigg( {g_{\sigma}\over{m_{\sigma}}} \bigg)^{-2}
( g_{\sigma} \sigma )^2    \nonumber \\
 & - & {1 \over{3}} b m_n (g_{\sigma} \sigma)^3
- {1 \over{4}} c (g_{\sigma} \sigma)^4
+ {1\over{2}}  \bigg({g_{\rho}\over{m_{\rho}}} \bigg)^2
\rho_{I_3}^{'2} ,
\end{eqnarray}
\begin{eqnarray}
\rho^H & = & \sum_{i=B,L}{ \rho_i }
+ {1\over{2}} \bigg({g_{\omega}\over{m_{\omega}}} \bigg)^2 \rho_{B}^{'2}
+ {1\over{2}} \bigg( {g_{\sigma}\over{m_{\sigma}}} \bigg)^{-2}
(g_{\sigma} \sigma)^2     \nonumber \\
& + & {1 \over{3}} b m_n (g_{\sigma} \sigma)^3
+ {1 \over{4}} c (g_{\sigma} \sigma)^4
+ {1\over{2}}  \bigg( {g_{\rho}\over{m_{\rho}}} \bigg)^2
\rho_{I_3}^{'2} .
\end{eqnarray}
Here $P_i$  and $\rho_i$  are the  expressions for  a Fermi gas of relativistic, non-interacting particles:
\begin{equation}
P_i =  {1 \over{3}} {\gamma_{i} \over{(2 \pi)^3} }
\int { d^{3}p
\;    { {p^2} \over{(p^2 + m_i^{*2})^{1/2}} }
\;     ( f_i(T) +  \bar{f}_i(T) )} ,
\label{Pres}
\end{equation}
\begin{equation}
\rho_i =  {\gamma_{i} \over{(2 \pi)^3} }
\int { d^{3}p
\;    {(p^2 + m_i^{*2})^{1/2}}
\;    ( f_i(T) +  \bar{f}_i(T) )} ,
\label{Ener}
\end{equation}
where  $f_i(T)$  and  $\bar{f}_i(T)$   are  the  Fermi$-$Dirac
distribution   functions   for   particles   and    antiparticles
respectively:
\begin{equation}
f_i(T) =
(\exp ( [ (p^2 + m_i^{*2})^{1/2} - \mu_i^* ] / T )  + 1 )^{-1} ,
\end{equation}
\begin{equation}
\bar{f}_i(T) =
(\exp ( [ (p^2 + m_i^{*2})^{1/2} + \mu_i^* ] / T )  + 1 )^{-1} .
\end{equation}
Note that for  baryons we use, instead of  masses $m_i$ and chemical potentials $\mu_i$,  ``effective'' masses $m_i^{*}$  and
chemical potentials $\mu_i^*$ given by:
\begin{equation}
m_i^{*} = m_i - x_{\sigma i} (g_{\sigma} \sigma) ,
\end{equation}
\begin{equation}
\mu_i^* = \mu_i
- x_{\omega i} \bigg({g_{\omega}\over{m_{\omega}}} \bigg)^2 \rho^{'}_{B}
- x_{\rho i} I_{3 i} \bigg( {g_{\rho}\over{m_{\rho}}} \bigg)^2
\rho^{'}_{I_3} .
\end{equation}

The hadronic  phase is  assumed to  be charge  neutral and  in chemical equilibrium under weak interactions.
Electric charge neutrality states:
\begin{equation}
n_p + n_{\Sigma^{+}} -  n_{\Sigma^{-}} - n_{\Xi^{-}}  - n_{e} = 0 .
\end{equation}
Chemical weak  equilibrium in  the presence  of trapped  electron
neutrinos implies that the chemical potential $\mu_i$  of each
baryon in the hadronic phase is given by:
\begin{equation}
\mu_i =  q_B \mu_n - q_e (\mu_e - \mu_{\nu_e} ) ,\label{potencial}
\end{equation}
where $q_B$  is its  baryon charge  and $q_e$  is its  electric
charge.    For  simplicity  we are assuming that  muon and tau
neutrinos are not present in the system, and their chemical potentials are  set
to zero.

The values of the five constants of the model  are determined by the properties of nuclear matter.
Three of them determine the nucleon couplings to the scalar, vector and vector-isovector mesons
$g_{\sigma}/m_{\sigma}$, $g_{\omega}/m_{\omega}$,
$g_{\rho}/m_{\rho}$, and the other  two determine the scalar self
interactions $b$ and $c$. Moreover, we consider that all hyperons in the
octet have the same coupling as the $\Lambda$, and that the coupling of the $\Lambda$ is 90 \% of that of the nucleons.
Therefore, we have that $x_{\sigma B}$, $x_{\omega B}$ and $x_{\rho B}$ are equal to $1$  for the nucleons and  0.9 for
 hyperons. In this paper we use the parametrization labelled as GM4 in \cite{Lugones_Grunfeld2011}
 with the following  values:  $\left({g_{\sigma}}/{m_{\sigma}}\right)^{2} = 11.79$ fm$^{2}$, $\left({g_{\omega}}/{m_{\omega}}\right)^{2} = 7.149$ fm$^{2}$, $\left({g_{\rho}}/{m_{\rho}}\right)^{2} = 4.411$ fm$^{2}$,
$b = 0.002947$, $c = -0.001070$. With this parametrization the EoS is stiff and gives a maximum mass of 2.0 $M_{\odot}$ for compact stars.

All the above equations can be solved numerically by  specifying
three thermodynamic quantities, e.g. the temperature $T$, the mass-energy density $\rho^H$
and the chemical potential of electron neutrinos in the hadronic phase $\mu_{\nu_e}^H$.

\section{Quark matter}\label{Q}
%
\subsection{Formalism for quark matter in bulk}
The quark phase is composed by \textit{u}, \textit{d}, and \textit{s} quarks, electrons, electron neutrinos and the corresponding antiparticles. We describe this phase using the MIT bag model at finite temperature with zero strong coupling constant, $m_{u,d} = 0$ and strange quark mass $m_{s} = 150$ MeV. The total thermodynamic potential for the quark matter phase can be written as 
\begin{equation}
\Omega_{Q} = \sum_{c, f} \Omega_{cf} +  \sum_{L} \Omega_{L} +  B ,
\label{omega}
\end{equation}
where $f  = u, d, s$ is the flavor index, $c = r, g, b$ is the color index and $L$ stands for the leptons. The contribution for free unpaired quarks is given by
\begin{eqnarray}
\Omega_{cf} = - \frac{\gamma T}{2 \pi^{2}}\int^{\infty}_{0}k^{2}\ln\left[1 + e^{-\left(\frac{E_{cf}-\mu_{cf}}{T}\right)}\right]dk, \label{omega_q}
\end{eqnarray}
being $E_{cf} = \sqrt{k^{2}+ m^{2}_{cf}}$ the particle energy, and $\mu_{cf}$ the chemical potential.
In the case of paired quarks we consider \cite{Schmitt_2010}
\begin{eqnarray}
\Omega_{cf} = - \frac{\gamma T}{2 \pi^{2}}\int^{\infty}_{0}k^{2}\ln\left[1 + e^{-\frac{\varepsilon_{cf}}{T}}\right]dk, \label{omega_delta}
\end{eqnarray}
where $\varepsilon_{cf} = \pm \sqrt{(E_{cf}-\mu_{cf})^{2}+ \Delta^{2}}$ is the single-particle energy dispersion relation  when it acquires an energy gap $\Delta$.
Note that, for particles, we can obtain Eq. (\ref{omega_q}) from Eq. (\ref{omega_delta}) in the limit $\Delta = 0$, by taking the minus sign in the dispersion relation for $E_{cf} < \mu_{cf}$ and the plus sign when $E_{cf} > \mu_{cf}$ (see e.g. \cite{Schmitt_2010}).

The gap equations for a color-superconducting condensate
with total spin $J=0$ have been derived perturbatively in dense QCD \cite{Pisarski1999}.
To leading order in weak coupling, the temperature dependence of the condensate is
identical to that in BCS-like theories \cite{Pisarski1999}. Then, we consider the following temperature dependence for the gap parameter in Eq. (\ref{omega_delta})
\begin{eqnarray}
\Delta(T) = \Delta_{0}\sqrt{1- \left(\frac{T}{T_{c}}\right)^{2},} \label{delta(T)}
\end{eqnarray}
where the critical temperature is $T_{c} = 0.57 \Delta_{0}$  \cite{Pisarski1999}.

Each lepton species contributes with a term of the form
\begin{eqnarray}
\Omega_{L} =  - \frac{\gamma T}{2 \pi^{2}}\int^{\infty}_{0}k^{2}\ln\left[1 + e^{-\left(\frac{E_L -\mu_L }{T}\right)}\right]dk, \label{omega_l}
\end{eqnarray}
with $L = e^-, e^+, \nu_e, \bar{\nu}_e $  and $E_L = \sqrt{k^{2}+ m_L^{2}}$. The degeneracy factor is $\gamma = 2, 2, 1$ for quarks, electrons and neutrinos, respectively. In all cases, the antiparticles contribution is obtained considering $\bar{\mu} = - \mu$.

In the present model, the quantities $B$, $m_s$ and {{$\Delta_0$}} are free parameters. According to their values, the energy per baryon of three flavor deconfined matter (composed of quarks \textit{u}, \textit{d} and \textit{s}) at zero pressure and temperature can be higher or lower than the mass of the neutron $m_{n}$. This condition defines the so called stability windows \cite{FarhiJaffe1984,Lugones2002}, which are the regions in the $m_s - B$ parameter space where quark matter is self-bound, i.e. it is the true ground state of strongly interacting matter.  In this paper we use $B=100$ MeV /fm$^{3}$ corresponding to absolutely stable quark matter and $B [ \mathrm{MeV /fm}^{3}] = 160, 353$ corresponding to quark matter {{allowed}} only at high pressures. The case $B$ = 353 MeV/fm$^{3}$ leads to an equation of state rather similar to \textit{set 1} of the NJL model used in Refs. \cite{Lugones_CarmoNJL, Lugones_Grunfeld2011}.

The here-considered \textit{just deconfined} phase is out of chemical equilibrium with respect to weak interactions. As we will see in the next section, the chemical potentials $\mu_{cf}$, $\mu_{e}$ and $\mu_{\nu_{e}}$  are related {{among each other}} through flavour conservation, {{color neutrality and pairing}} conditions.

\subsection{Finite size effects}

In order to study the formation of finite size droplets of quark matter we use the multiple reflection expansion (MRE) formalism  \cite{Lugones_Grunfeld2011,Balian1970, Madsen, Kiriyama2003, Kiriyama2005}. {{For a finite spherical droplet}}, the modified density of states  is given by
\begin{eqnarray}
\rho_{{MRE}}(k, m_f, R) =  1 + \frac{6\pi^2}{k R} f_S +  \frac{12\pi^2}{(k R)^2}f_C 
\label{rhoMRE}
\end{eqnarray}
with
\begin{eqnarray}
f_S  &=& -\frac{1}{8 \pi}\bigg[1 - \frac{2}{\pi}\arctan \bigg(\frac{k}{m_f}\bigg)\bigg], \label{fS} \\
f_C  &=& \frac{1}{12 \pi^2}\bigg[1 - \frac{3k}{2m_f}\bigg(\frac{\pi}{2}- \arctan \bigg(\frac{k}{m_f}\bigg)\bigg)\bigg] .
\label{fC}
\end{eqnarray}
The $f_S$ and $f_C$ terms correspond to the contributions of surface and curvature, respectively, and the drop radius is given by $R$. As in \cite{Lugones_Grunfeld2011}, we use the Madsen ansatz \cite{Madsen} to take into account the finite quark mass contributions.

The MRE formalism is included in the equation of state by means of the following replacement in the thermodynamic integrals
\begin{eqnarray}
\int_0^{\infty} d^3 k \rightarrow \int_{\Lambda_{IR}}^{\infty} d^3 k \quad \rho_{{MRE}}.
\label{troca}
\end{eqnarray}
According to  Ref. \cite{Kiriyama2003, Kiriyama2005}, the density of states for massive quarks is reduced if compared with the bulk density, leading to a negative density of states for a range of small momentum. In order to avoid this nonphysical effect one can introduce an infrared cutoff $\Lambda_{IR}$ in momentum space $k$ \cite{Kiriyama2003, Kiriyama2005}. For calculating the value of $\Lambda_{IR}$, we need to solve the equation $\rho_{MRE} = 0$ with respect to the momentum $k$ and take the larger root as the IR cutoff. Then, this cutoff depends as well on the radius of the spherical drop and the quark mass.

Performing the above replacement into Eq. (\ref{omega_q}), the thermodynamic potential for the quark matter phase reads
\begin{equation}
\Omega_{Q}^{{MRE}} = - P^{Q} V^{Q} + \sigma A + \zeta C
\end{equation}
Here, $ A = 4\pi R^2$ is the spherical drop area, $C = 8 \pi R$ the curvature and {{ $V^{Q} = \frac{4}{3} \pi R^3$ }} the volume.
The pressure $P^{Q}$ is given by
\begin{eqnarray}
 {{  P^{Q} \equiv -\frac{\partial \Omega_{Q}^{{MRE}}}{\partial V^{Q}} \bigg|_{T, \mu, A, C} \; , }}
\end{eqnarray}
the surface tension is
\begin{eqnarray}
 {{  \sigma \equiv \frac{\partial \Omega_{Q}^{{MRE}}}{\partial A} \bigg|_{T, \mu, V^{Q}, C}        }}
\end{eqnarray}
and the curvature energy density is obtained as follows
\begin{eqnarray}
 {{  \zeta \equiv \frac{\partial \Omega_{Q}^{{MRE}}}{\partial C} \bigg|_{T, \mu, A, V^{Q}} \; .    }}
\end{eqnarray}
%

\section{Nucleation of quark droplets in hadronic matter}

To calculate the conditions for the transition  we assume  thermal, chemical and mechanical equilibrium between the quark droplet and its hadronic environment. Thermal and chemical equilibrium means that the temperature $T$ and the Gibbs free energy per baryon $g$ are equal in both the hadronic (H) and the quark matter (Q) phases:
\begin{eqnarray}
T^{H}  &=&  T^{Q} \;  \equiv \; T ,  \label{dec01}  \\
g^{H}(T, {{\rho^H}}, \mu_{\nu_{e}}^{H})   &=&  g^{Q}(T, \{ \mu_{fc} \}, \mu_{e}^{Q}, \mu_{\nu_{e}}^{Q}).
\label{dec02}
\end{eqnarray}
To obtain the condition for mechanical equilibrium we write the total thermodynamic potential for the combination of a just nucleated quark matter drop immersed in a homogeneous hadronic environment: $\Omega = \Omega_{H} + \Omega_{Q}^{{MRE}}$,  where  $\Omega_{H} =  - P^{H} V^{H}$ for the hadronic phase. The condition for mechanical equilibrium is obtained from $\partial \Omega / \partial R = 0 $ and reads \cite{Lugones_Grunfeld2011,Landau}:
\begin{eqnarray}
P^{Q}(T, \{ \mu_{fc} \},  \mu_{e}^{Q}, \mu_{\nu_{e}}^{Q})  -  P^{H}(T, {{\rho^H}}, \mu_{\nu_{e}}^{H}) =\nonumber \\ 
\frac{2}{R} \sigma(T, \{ \mu_{fc} \},  \mu_{e}^{Q}, \mu_{\nu_{e}}^{Q})   +  \frac{2}{R^2} \zeta(T, \{ \mu_{fc} \},  \mu_{e}^{Q}, \mu_{\nu_{e}}^{Q}).
\label{mechanical_eq}
\end{eqnarray}

On the other hand, deconfinement is driven by strong interactions and therefore quark and lepton flavours must be conserved during the deconfinement transition \cite{IidaSato1998,Lugones2005,Lugones9899,Madsen1994,Bombaci2004,Bombaci2007}.
When a small quark-matter drop is nucleated at the core of a compact star, the abundances of all particle species inside it must be \textit{initially} the same as in the hadronic matter from which it has been originated. Thus we have
\begin{eqnarray}
Y_{f}^{H}(T, {{\rho^H}}, \mu_{\nu_{e}}^{H})  =  Y_{f}^{Q}(T, \{ \mu_{fc} \}, \mu_{e}^{Q}, \mu_{\nu_{e}}^{Q}),  \label{dec03}
\end{eqnarray}
with $f = u, d, s, e, \nu_e$, being  $Y^H_f \equiv n^H_f / n^H_B$ and  $Y^{Q}_f \equiv n^{Q}_f /
n^{Q}_B$ the abundances of each particle species in the hadronic and quark
phase respectively. Notice that, since the hadronic phase is assumed to be
electrically neutral, flavor conservation ensures automatically
the charge neutrality of the just deconfined quark phase. As a consequence of Eq. (\ref{dec03}), in the present work,  the just deconfined phase is out of $\beta$ equilibrium.

Additionally, the deconfined phase must be locally colorless, i.e. it must be composed by an equal
number or \textit{red}, \textit{green} and \textit{blue} quarks:
\begin{eqnarray}
n_{r}(T, \{ \mu_{fr} \})  = n_{g}(T, \{ \mu_{fg} \}) , \label{dec04}\\
n_{r}(T, \{ \mu_{fr} \})  = n_{b}(T, \{ \mu_{fb} \}) . \label{dec05}
\end{eqnarray}

Finally, it has been shown that when color superconductivity is included together with flavor
conservation and color neutrality, the most likely configuration of the just deconfined
phase is 2SC provided the pairing gap is large enough \cite{Lugones2005}.
Thus, in order to allow for pairing between quarks $d_{r}$ with $u_{g}$ and between quarks $u_{r}$ with $d_{g}$
we impose that:
\begin{eqnarray}
n_{ur}(T, \mu_{ur}) = n_{dg}(T, \mu_{dg}) , \label{dec06} \\
n_{dr}(T, \mu_{dr}) = n_{ug}(T, \mu_{ug}). \label{dec07}
\end{eqnarray}
Equations (\ref{dec01}-\ref{dec07}) together with the assumption that $\mu_{sr} = \mu_{sg} = \mu_{sb}$
allow to determine the transition density for  given values of the droplet radius, the temperature and
the chemical potential of the neutrinos in the hadronic phase.

The theory of homogeneous nucleation predicts a rate of nucleation of a droplet of radius $R$ as  given by \cite{Madsen,Landau}
\begin{eqnarray}
\Gamma \approx T^4 e^{-\Delta \Omega / T },
\label{taxa_nucl}
\end{eqnarray}
where $\Delta\Omega$, the free energy involved in the formation of the droplet, is given by
\begin{eqnarray}
\Delta\Omega =  - \frac{4}{3} \pi R^3 (P^Q - P^H) + 4\pi R^2 \sigma + 8 \pi R \zeta.
\label{deltaOmega}
\end{eqnarray}

The prefactor $T^4$  is included in Eq. (\ref{taxa_nucl}) on dimensional grounds because the nucleation rate is largely dominated by the exponential \cite{Madsen1994,Lugones_Grunfeld2011,Horvath1992,Horvath1994}, i.e. we always have $\log_{10} \Gamma \approx \log_{10} (\mathrm{prefactor})  -  \Delta \Omega/[ T  \ln(10)]$ with the second term much larger than the first. A more elaborate statistical prefactor has been developed  in the literature \cite{prefactor}. However, the determination of such prefactor involves the specification of transport coefficients such as the thermal conductivity, and the shear and bulk viscosities, which are not well known for ultradense matter (see Refs. \cite{Logoteta,prefactor} for more details). We have checked that the prefactor given in Ref. \cite{Logoteta} is in fact very different from the  $T^4$ factor, but it is not dominant with respect to the exponential for the conditions encountered in our calculations. Thus, our results are not significantly affected by the choice in Eq. (\ref{taxa_nucl}). The nucleation time $\tau$ is the typical time needed to nucleate a droplet of radius $R$ and is given by $\tau =( \frac{4}{3} \pi R^3 \ \Gamma )^{-1}$.

\section{Results}
%
\begin{figure}[tb]
\begin{flushright}
\includegraphics[scale=0.37]{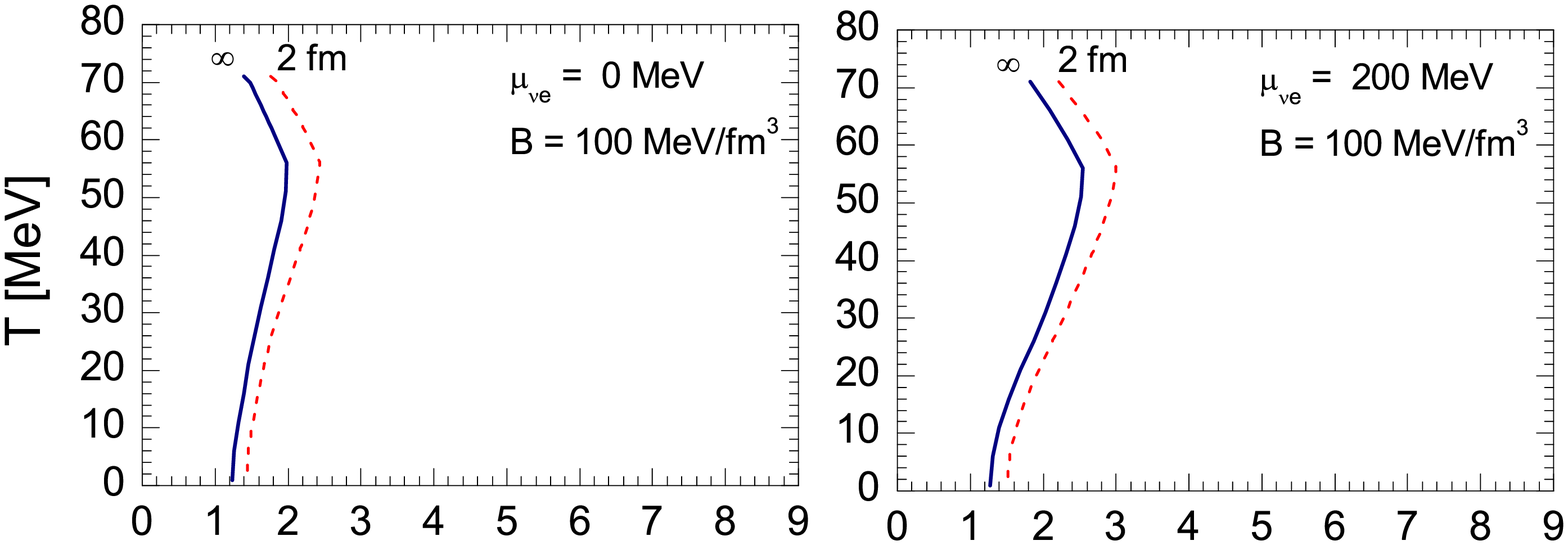}\\
\includegraphics[scale=0.37]{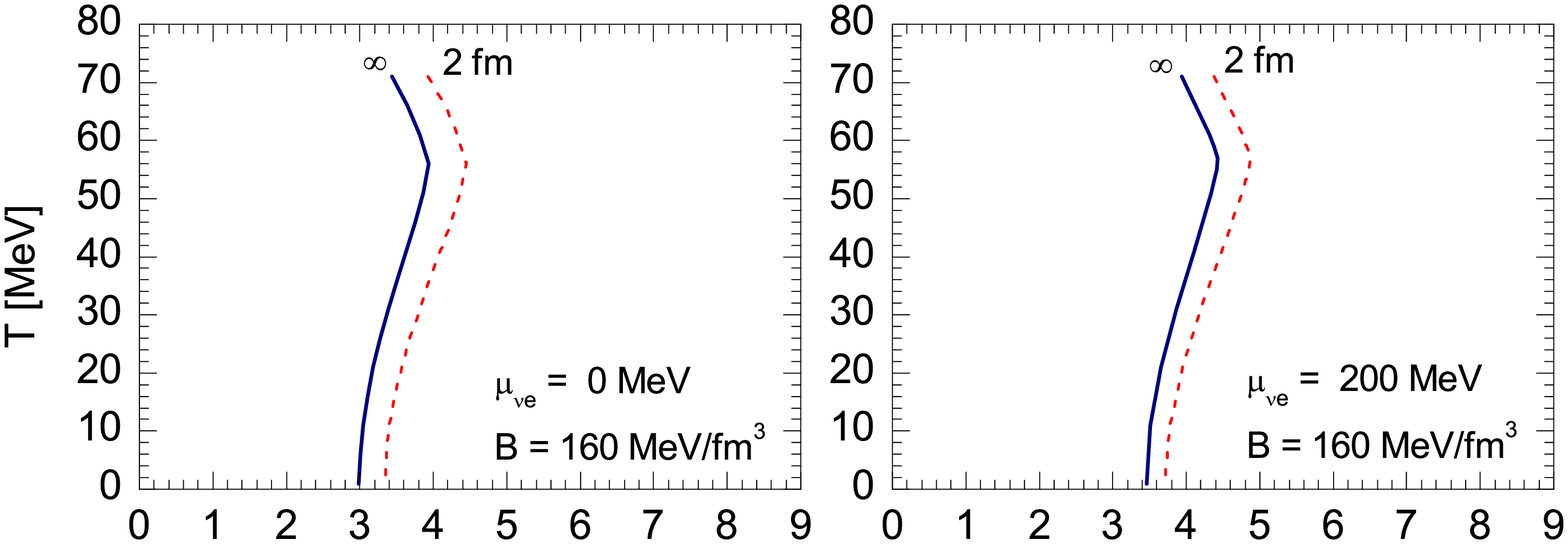}\\
\includegraphics[scale=0.37]{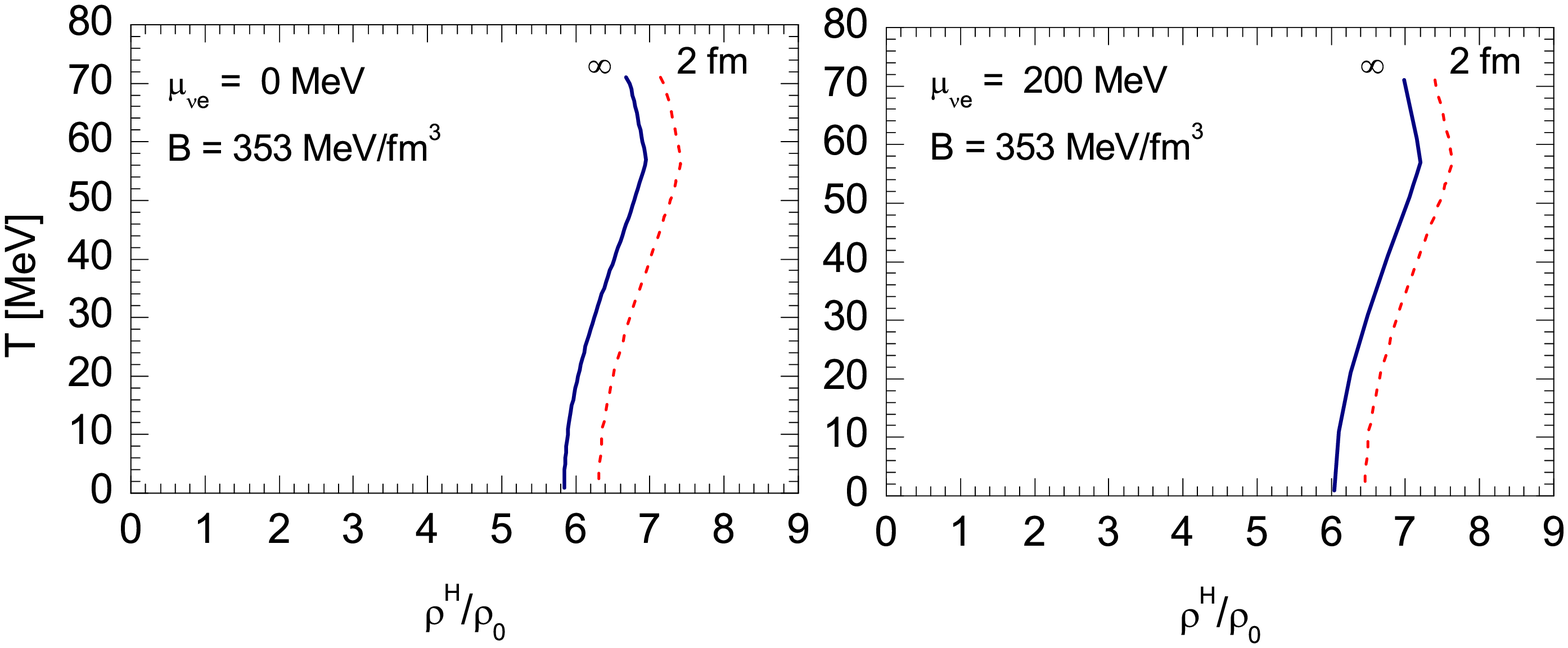}
\end{flushright}
\caption{Mass-energy density and temperature of hadronic matter at which deconfinement is energetically favoured for $R= \infty$ and $R= 2$ fm. The deconfinement density for finite size drops is larger than for the bulk case due to surface tension and curvature energy. We have calculated the transition density for several radii and found that for $R \sim 100-200$ fm the curves are almost coincident with the case $R= \infty$. The density is given in units of the nuclear saturation density $\rho_{0}$. We use $\Delta_{0}$ = 100 MeV  and  B [MeV/fm$^{3}$] = 100, 160 and 353. We considered different chemical potentials for trapped electron neutrinos in hadron matter {{but for simplicity we show only the results for}}  $\mu_{\nu_{e}}^{H} $ [MeV] =  0 and 200. }
\label{1}
\end{figure}
%
In Fig. \ref{1} we show the mass-energy density and the temperature at which the conversion of a portion of hadronic matter into a  quark droplet is energetically favourable. By comparing {{left and right}} panels, we can see that curves with $\mu_{\nu_{e}}^{H} = 200$ MeV are displaced to the right with respect to the same curves for $\mu_{\nu_{e}}^{H} = 0$ MeV. In other words, the trapped neutrinos push the {{transition density}}  to higher values, i.e. the trapped neutrinos tend to inhibit the transition, although the effect is not very large. This behaviour is also reported in Ref. \cite{Lugones9899}.  Another interesting feature of the curves shown in Fig. \ref{1} is that  for $T < T_c \approx 57$ MeV there is a significant decrease of the transition density, which is associated with the appearance of  color superconductivity below $T_c$.  This results are in agreement with similar calculations that employ the Nambu-Jona-Lasinio model for quark matter \cite{Lugones_Grunfeld2011}. As expected, we found that the  transition density increases when the drop radius decreases, {{as a consequence of the larger contribution of the surface and curvature terms in Eq. (\ref{mechanical_eq})}}.

\begin{figure}[tb]
\begin{flushright}
\includegraphics[scale=0.4]{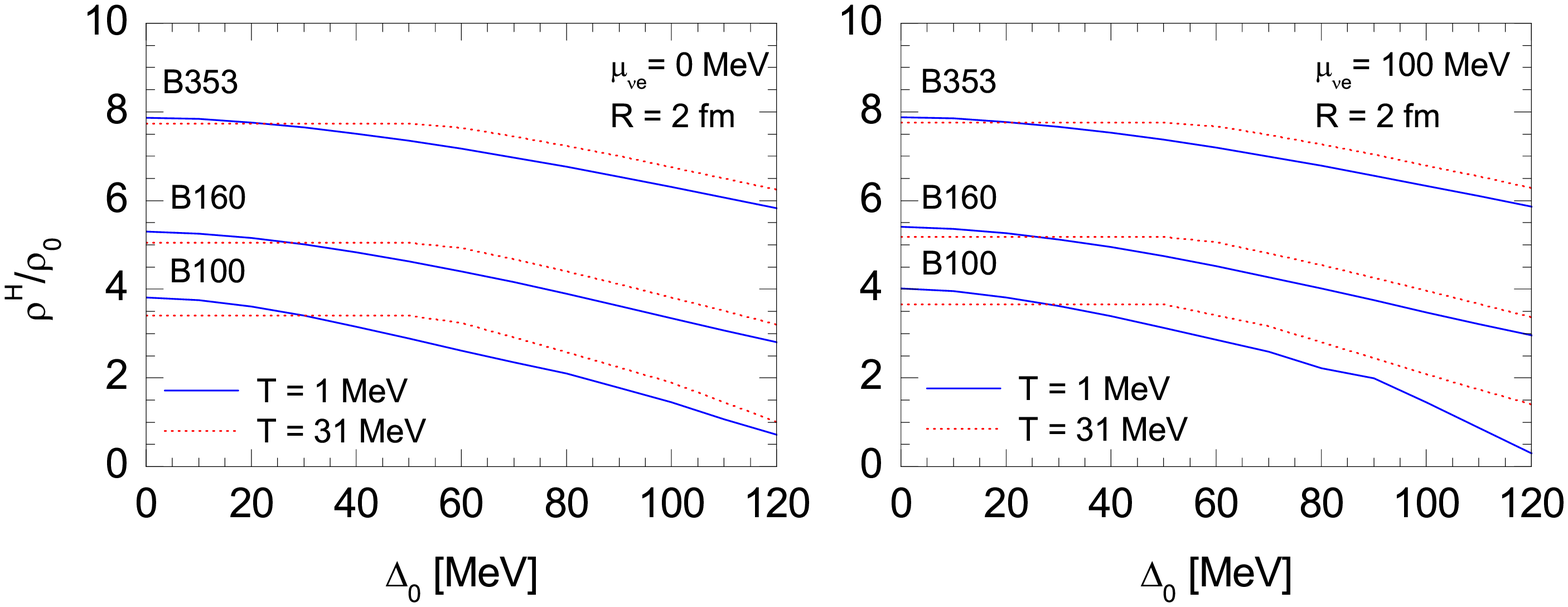}\\
\end{flushright}
\caption{Mass-energy density at which deconfinement is energetically favoured as a function of the parameter $\Delta_0$.
There is a strong decrease of the transition density $\rho^H/\rho_0$ for large enough $\Delta_0$.}
\label{2}
\end{figure}
%
%
The behaviour of the transition density with the gap parameter $\Delta_0$ is shown in Fig. \ref{2} for different values of $\mu_{\nu_{e}}^{H}$, $T$ and $B$.  Notice that the  transition density is a decreasing function of the gap parameter $\Delta_{0}$.
The effect is strong, e.g. the transition density for $\Delta_0 \sim 100$ MeV is much smaller than for $\Delta_0$ = 0 MeV.
For sufficiently small $\Delta_{0}$ the transition density has constant values. This is because this part of the curve corresponds to temperatures that are larger than the critical temperature $T_{c} = 0.57 \Delta_{0}$, and therefore the pairing gap $\Delta(T)$ is zero.

\begin{figure}[tb]
\begin{flushright}
\includegraphics[scale=0.4]{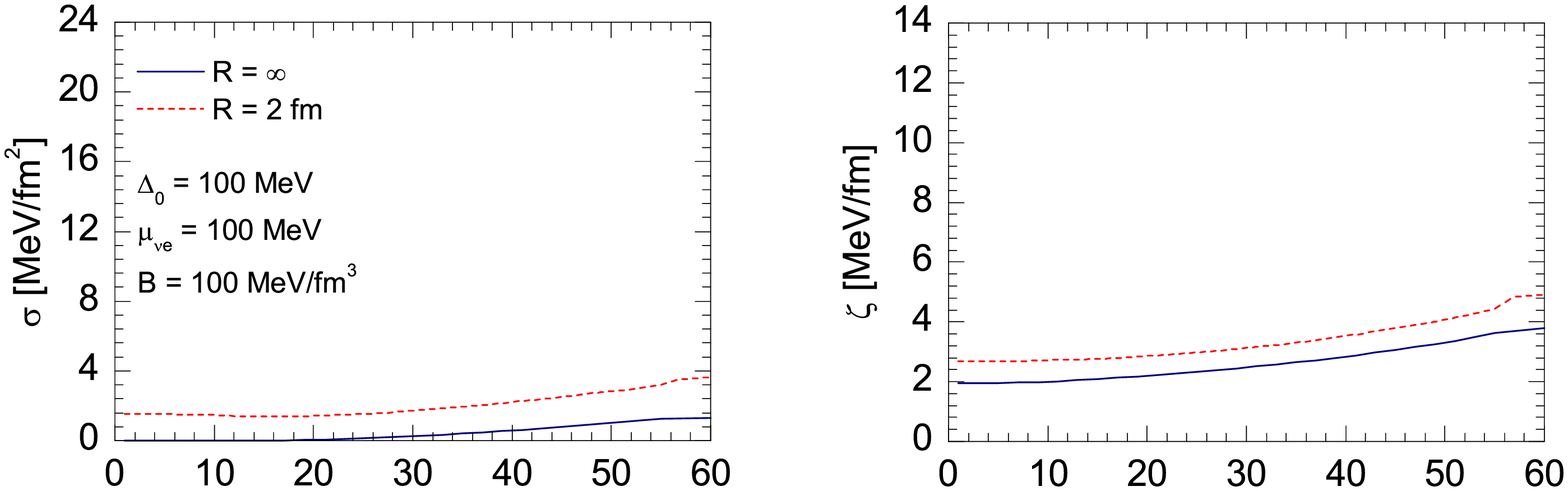}\\
\includegraphics[scale=0.4]{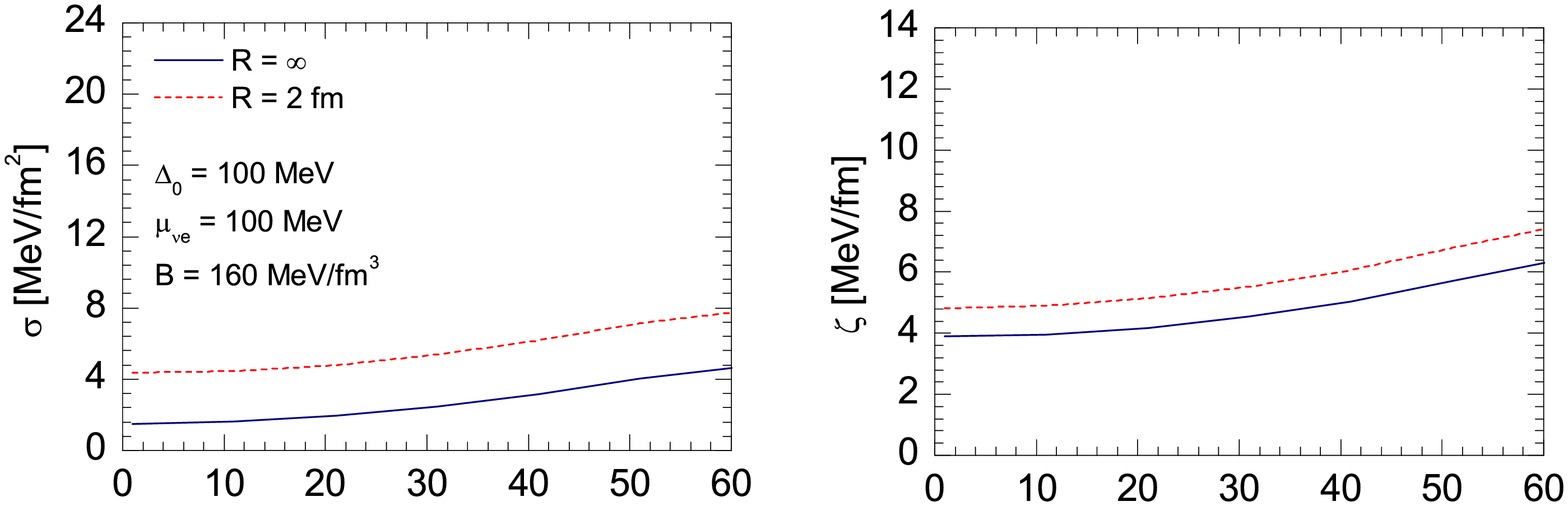}\\
\includegraphics[scale=0.4]{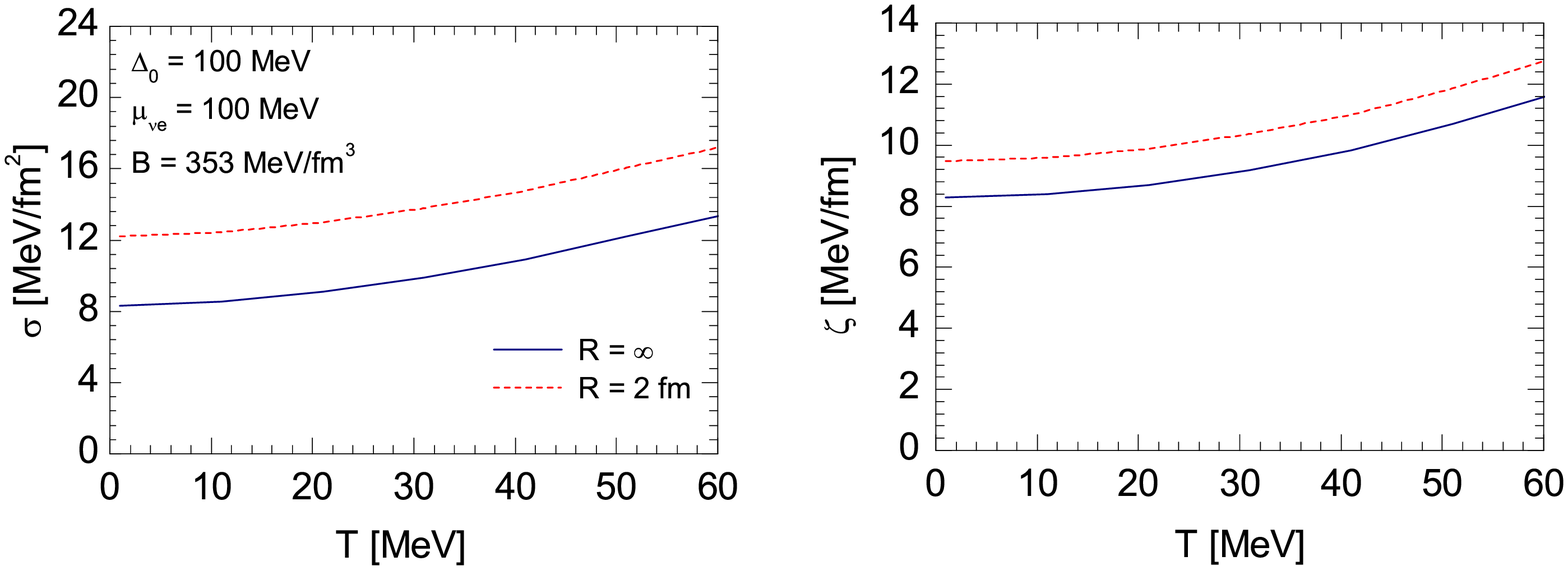}
\end{flushright}
\caption{Surface tension $\sigma$  and curvature energy $\zeta$ for droplets with $R= 2$ fm and $R=\infty$. The curves were calculated  using the set of values of the thermodynamic variables that arise from Eqs. (\ref{dec01}$-$\ref{dec07}), i.e. the same set of values that led to Fig. 1.       }
\label{3}
\end{figure}
%
As can be seen in Fig. \ref{3}, we obtain   {{$\sigma \lesssim 18 \ \mathrm{MeV \  fm}^{-2}$}} and $\zeta \sim 4-12 \ \mathrm{MeV \ fm}^{-1}$ which are {{ordinarily}} an order of magnitude smaller than the values obtained within the Nambu-Jona-Lasinio model \cite{Lugones_Grunfeld2011}. {{Our values for $\sigma$ are also smaller than the constant value $\sigma = 30\ \mathrm{MeV \  fm}^{-2}$ assumed in Ref. \cite{Logoteta}.  }}

\begin{figure}[tb]
\begin{flushright}
\includegraphics[scale=0.4]{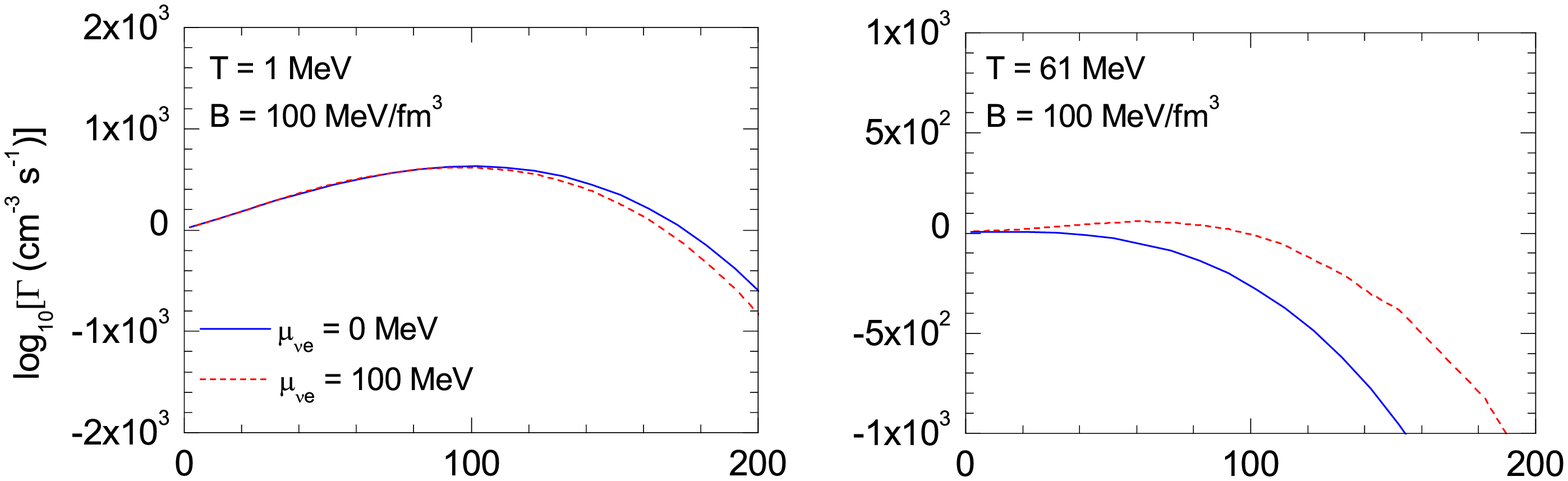}\\
\includegraphics[scale=0.4]{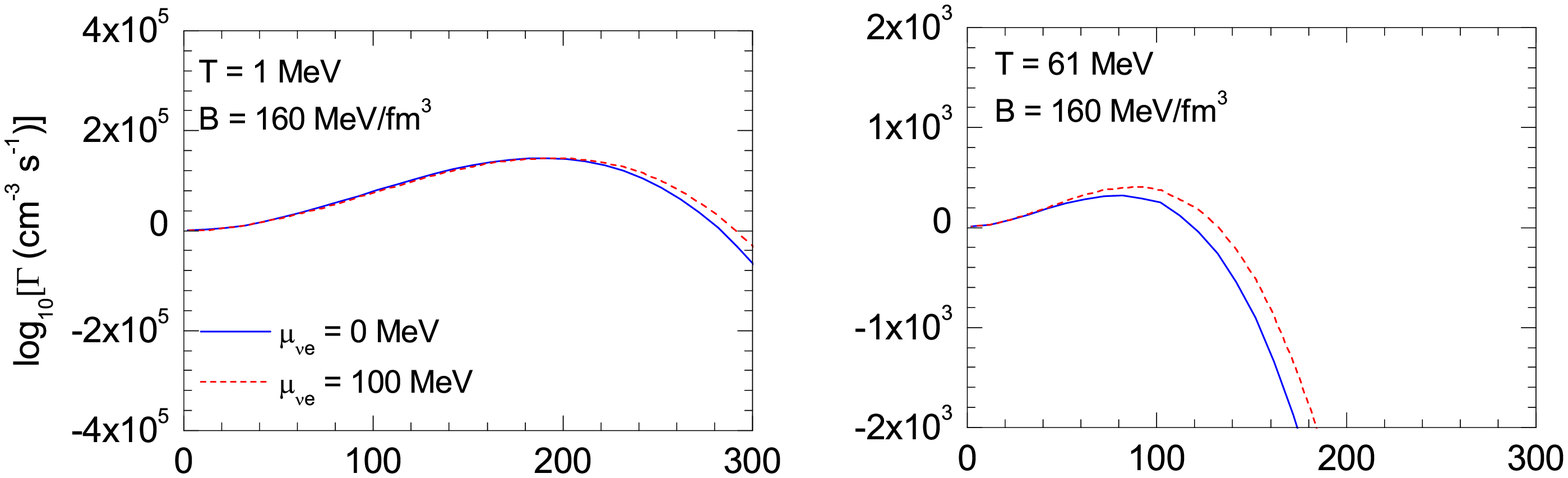}\\
\includegraphics[scale=0.4]{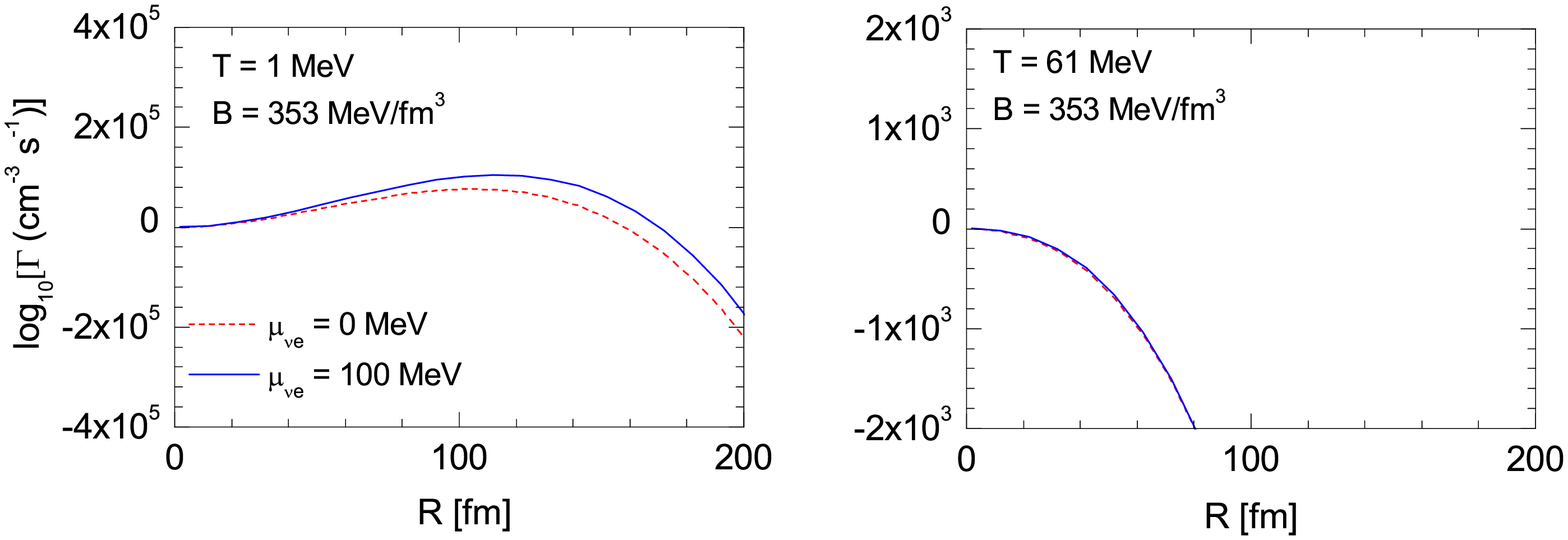}
\end{flushright}
\caption{Nucleation rate of droplets as a function of their radii, for $\Delta_0 = 100$ MeV, B [MeV/fm$^{3}$] = 100, 160, 353,  $\mu_{\nu_{e}}^{H}$ [MeV] = 0, 100 and T [MeV] = 1, 61.  }
\label{4}
\end{figure}

\begin{figure}[tb]
\begin{flushright}
\includegraphics[scale=0.4]{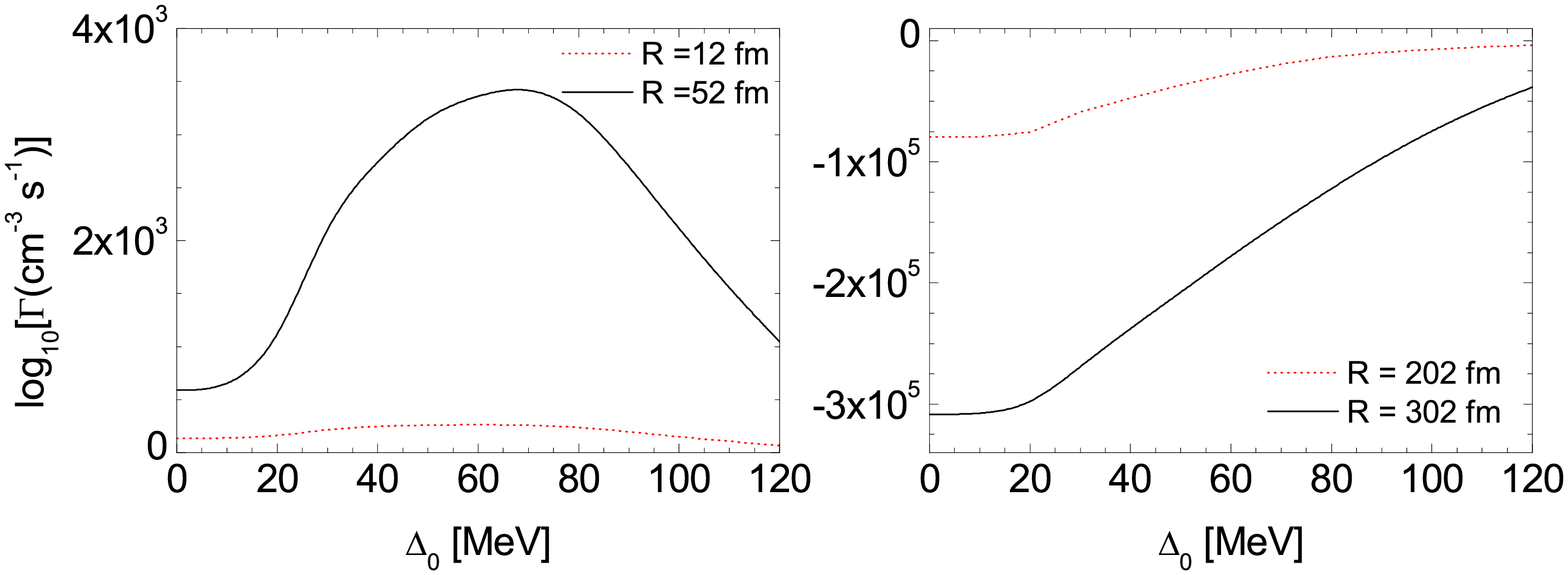}
\end{flushright}
\caption{Nucleation rate of droplets as a function of  $\Delta_0$  for B [MeV/fm$^{3}$] =  160,  $\mu_{\nu_{e}}^{H}$ [MeV] = 100 and T [MeV] = 11. }
\label{5}
\end{figure}

In Figs. \ref{4} and \ref{5}  we present the results for the nucleation rate given by Eq. (\ref{taxa_nucl}). From Fig. \ref{4} we see that there are large variations in the value of $\log_{10}(\Gamma [ \mathrm{cm}^{-3} \mathrm{s}^{-1}])$ for different values of $B$, but the qualitative behaviour is nearly the same in all cases:  (i) the nucleation rate increases hugely  when the temperature falls from 61 MeV to 1 MeV, and (ii) the nucleation of droplets larger than some value $R^*$  are strongly suppressed (typically $R^* \sim 100-200$ fm). On the contrary, droplets with $R \lesssim R^*$ have a huge nucleation rate at low enough $T$, and in practice should nucleate instantaneously when the bulk transition density is attained at the core of a protoneutron star. In Fig. \ref{5} we show that the main conclusions obtained from Fig. \ref{4} are valid for different values of $\Delta_0$. 
Specifically, we notice on the right panel that $\log_{10} \Gamma$ is a large negative number for radii above $R^*$. On the left panel, we see that $\log_{10} \Gamma$ is a large positive number for radii below $R^*$, with the possible exception of some very small drops with $R \lesssim 10$ fm.

\section{Summary and Conclusions}

In the present work we analyse the nucleation of quark droplets inside a protoneutron star focusing our analysis in the color superconductivity, trapped neutrinos, and finite size effects. To describe strongly interacting matter we consider a different model for each phase. For the hadronic phase we use the Walecka model including the baryonic octet, electrons and electron neutrinos in equilibrium under weak interactions. The just-deconfined quark matter is described using the MIT bag model (considering $u$, $d$ and $s$ flavours, electrons and electron neutrinos). To determine the energy density at which the deconfinement occurs we assume a first order phase transition, we impose flavour conservation during the transition, and we consider the deconfined  phase to be in the energetically favoured  2SC state. The effects of finite size of the nucleated droplet are included through {{the}} MRE formalism, i.e. the surface tension and the curvature energy are not treated as free parameters but are calculated self-consistently as a function of the thermodynamic state of the system.

Firstly, we determine the transition density for  given values of the droplet radius, the gap parameter $\Delta_0$, the temperature and the chemical potential of the neutrinos in the hadronic phase (see Figs. \ref{1} and \ref{2}). Due to the effect of color superconductivity, there is a significant decrease of the transition density at low temperatures.  In most cases, our values for the surface tension (see Fig. \ref{3}) turn out to be a factor $\sim 2-10$ smaller that the constant value assumed in Ref. \cite{Logoteta} within the MIT bag model and between one and two orders of magnitude smaller than the value calculated in \cite{Lugones_Grunfeld2011}  using the MRE together with the NJL model.  
Secondly, we calculate the nucleation rate of quark droplets as a function of their radii for different temperatures, pairing gaps $\Delta_0$, and trapped neutrino abundances (see Figs. \ref{4} and \ref{5}). Large droplets have a low transition density because surface and curvature effects tend to be small; but at the same time, their nucleation probability is suppressed because there is a large free energy involved in their formation (due to their large volume). As a consequence,  the nucleation rate $\Gamma$ is negligible for droplets larger than  $R^* \sim 100-200$ fm. We also show that $\Gamma$ is huge for most radii below  $R^*$ provided that the temperature is low enough. 

All these results are in qualitative agreement with previous calculations obtained within the frame of the Nambu-Jona-Lasinio model \cite{Lugones_Grunfeld2011}. Quantitatively, the here obtained nucleation rate is very different to the one obtained within the NJL model; nonetheless, both are negligible above some radius $R^*$ and huge below it. Moreover,  in spite of the large differences in $\sigma$, $\zeta$ and $\Gamma$ for the MIT and the NJL models, the radius $R^*$ is within the same order of magnitude in both cases (some hundreds of fm).  
Since the transition density for $R \sim 100-200$ fm is almost coincident with the bulk transition density and the nucleation rate of such droplets is huge,  the deconfinement transition should be triggered instantaneously when a density slightly larger than the bulk transition density shown in Fig. 1 is reached at some layer of a PNS. Since color superconductivity lowers the transition density at low temperatures,  the transition is likely to occur after significant cooling in a massive enough protoneutron star.

\section*{Acknowledgements}  T. A. S. do Carmo acknowledges the financial support received from CAPES.
G. Lugones acknowledges the financial support received from FAPESP.

\end{document}